\documentclass[aps,twocolumn,groupedaddress,showpacs]{revtex4}
\usepackage{graphicx}%

\begin{document}
\bibliographystyle{apsrev}


\title{The core structure of presolar graphite onions}


\author{P. Fraundorf}
\email[]{pfraundorf@umsl.edu}
\author{Martin Wackenhut}

\affiliation{Physics \& Astronomy and Center for Molecular Electronics, 
U. Missouri-StL (63121) \\ St. Louis, MO, USA}


\date{\today}

\begin{abstract}
  Of the ``presolar particles'' extracted from carbonaceous meteorite dissolution residues, i.e. of those particles which show isotopic evidence of solidification in the neighborhood of other stars prior to the origin of our solar system, one subset has an interesting concentric graphite-rim/graphene-core structure.  We show here that single graphene sheet defects in the onion cores (e.g. cyclopentane loops) may be observable edge-on by HREM.  This could allow a closer look at models for their formation, and in particular strengthen the possibility that growth of these assemblages proceeds atom-by-atom with the aid of such in-plane defects, under conditions of growth (e.g. radiation fluxes or grain temperature) which discourage the graphite layering that dominates subsequent formation of the rim.
\end{abstract}
\pacs{97.10.Me, 98.38.Cp, 96.50.Mt, 61.46.+w}

\maketitle

\section{Introduction}

``High-density'' and ``isotopically-heavy'' presolar graphite onions from the Murchison meteorite \cite{Amari95} appear to have been created by the concentric growth of 0.34 nm graphite shells.  Often these {\em rims} appear to have nucleated not on the inorganic grains that Bernatowicz et al. \cite{Bernatowicz96} have used to constrain condensation conditions, but on a spherical {\em core} consisting of a novel carbon phase with density comparable to graphite.  This phase is comprised of atom-thick ``flakes'' typically less than 4 nm across, with graphite (hk0) ordering but with no sign of the 0.34 nm (002) graphite spacing characteristic of graphite, amorphous carbon, and multiwall carbon nanotubes, i.e. of most solid non-diamond carbon phases \cite{Bernatowicz95}.  Thus, as with the onions themselves, the onion cores may provide information on astrophysical environments (e.g. cool stellar outflows) that generate significant amounts of condensed carbon in the galaxy \cite{Jura97}. 

One question that arises is the extent to which cyclopentyl (or other layer defect) groups play a role in facilitating three-dimensional growth in core material whose periodicity seems to be dominated by the planar (cyclohexyl) ordering of two-dimensional graphene sheets \cite{Allamandola89}.  At one extreme, the cores might be formed by the ``collisional agglomeration'' of previously-formed planar polycyclic aromatic hydrocarbons (PAHs) \cite{Bernatowicz97c}.  At the other extreme, the cores might form by the addition of one carbon atom at a time, with layer defects added in randomly from time to time to lessen entropy loss during the condensation process \cite{Sedlmayr97}.  In other words, can the detailed structure of collected onion cores provide further insight into the creation of their disordered three-dimensional  form from components with well-defined two-dimensional nearest-neighbor order? 
 
\begin{figure}[tbp]
\includegraphics{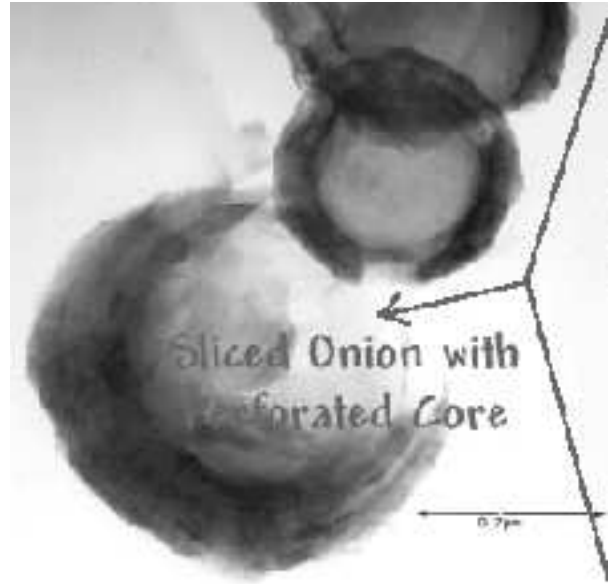}%
\caption{Image of three onion slices, showing one with a torn core suitable 
for electron phase contrast study.}
\label{Fig1}
\end{figure}

\section{Experimental Setup}

Interstellar graphite onions used for this study were from the Murchison meteorite graphite separate KFC1 (2.15-2.20 g/cm$^3$), whose preparation has been described in detail by Amari et al. \cite{Amari94}.  The onions were microtomed, and deposited onto 3mm copper grids with a carbon support film, by Bernatowicz et al. \cite{Bernatowicz96}.  Specimens were examined in a 300 kV Philips EM430ST TEM with point resolution near $0.2 nm$.  Because the sections were generally too thick for electron phase contrast imaging, sliced onions with torn regions of core located over holes in the underlying support film (c.f. Fig \ref{Fig1}) were our primary target.  Many of the microtome sections on the TEM grid designated KFC1A:E have been surveyed for HREM suitability and other properties (e.g. their core-rim ratio).  The images reported here come from onions on that mount.

Electron phase contrast (HREM) images were recorded on Kodak electron image film at 
240,000x or 700,000x, and then digitized at 2500 dots per inch and 12 bits per pixel with an Agfa DuoScan2500 scanner.  Image analyses and simulations were done in Semper, C, Visual Basic, Cerius 2, EMS, and Mathematica.

\section{Core-Structure}

HREM images of very thin specimens are {\em to first order} projections of electron column density in the specimen along the direction of the beam, suitably ``band-pass'' filtered by the complex contrast transfer function of the microscope \cite{Spence88}.  Hence individual graphene sheets will be difficult to view face-on unless the specimen is no more than a few atoms thick, given that the absence of graphite layer spacings in diffraction argues strongly that the other atoms in the specimen will not be periodically arrayed above or below.  

\begin{figure}[tbp]
\includegraphics{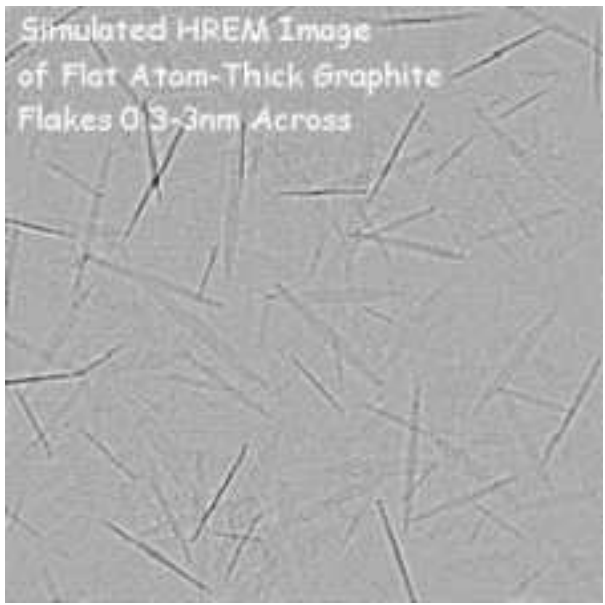}%
\caption{Strong phase-object simulation of 0.19 nm resolution HREM image from 
a collection of randomly-oriented atom-thick hexagonal graphene flakes.}
\label{Fig2}
\end{figure}

\begin{figure}[tbp]
\includegraphics{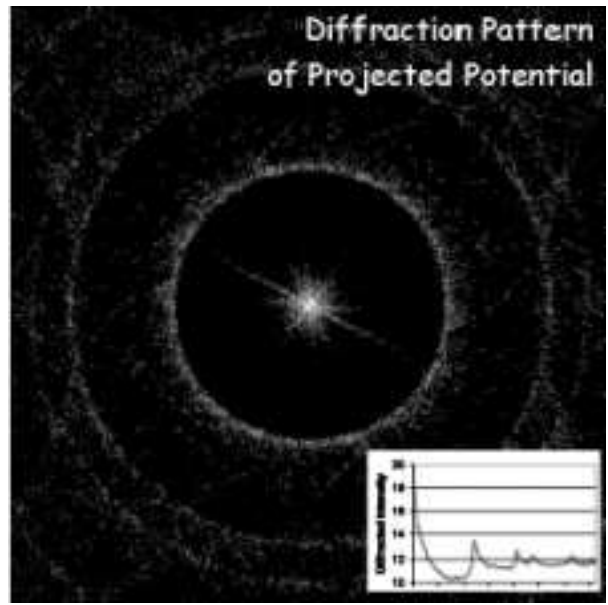}%
\caption{Simulated diffraction pattern of the previous image, showing both the spacings and asymmetric line broadening familiar from experimental electron diffraction patterns of onion cores.}
\label{Fig2b}
\end{figure}

The story is different for atom-thick sheets viewed edge-on.  In fact, the properties of lattice fringes from single planes embedded in an otherwise disordered field might be easier to recognize than those of lattice fringes in a crystal, since Fourier ringing from adjacent planes in the latter case may obscure single column defects.  This is borne out by image simulations like that shown in Figs. \ref{Fig2} and \ref{Fig2b}.  The visual signature of graphene sheets randomly distributed in an otherwise random matrix is therefore in large part a collection of randomly-oriented line segments.

\begin{figure}[tbp]
\includegraphics{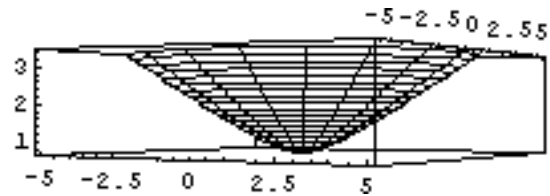}%
\caption{Hyperbolic cone with perimeter equal to 5/6 that of a circle at the same distance from the center.}
\label{Fig3b}
\end{figure}

One might imagine that a single cyclopentane embedded in a graphene sheet would impose only a gradual curvature effect, diluted by the otherwise large number of hexagons present therein.  On the contrary, the effect is quite large, beginning with bond angles for the 5-hexagon corannulene molecule bent up from the plane of the pentagon by $26.8^{\circ}$ \cite{Dresselhaus96}.  In the asymptotically large-molecule limit, the envelope may be approximated by a hyperbolic-cone with a half-angle of $56.44^{\circ}$, to take into account the 5/6 perimeter mismatch between this structure and that of a disk of the same radius (cf. Fig \ref{Fig3b}).  

\begin{figure}[tbp]
\includegraphics{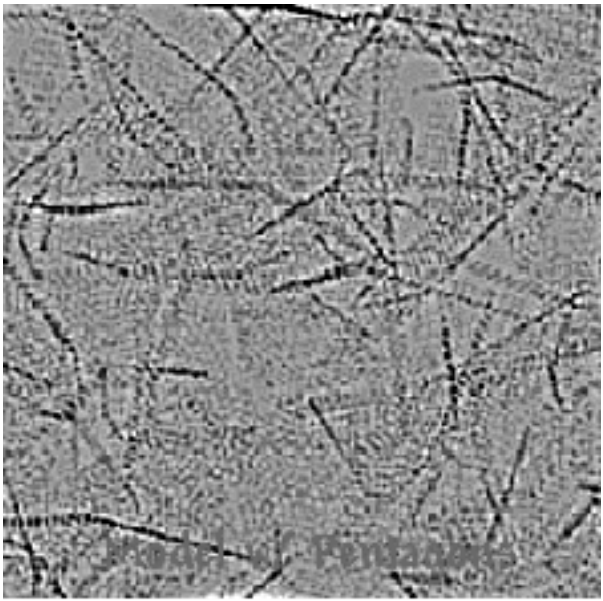}%
\caption{Strong phase-object simulation of 0.19 nm resolution HREM image from 
a collection of randomly-oriented atom-thick hexagonal graphene flakes with a 
cyclopentane at their center.}
\label{Fig3}
\end{figure}

When a simple model of five graphene ``flats'' intersecting at a central cyclopentane is randomly distributed at near-graphite densities into a 100 nm cube, and then sliced for HREM image simulation, the result is shown in Fig. \ref{Fig3}.  As you can see, the visual signature is subtly perturbed by the addition of line segments that make small angles with one another, and that intersect in projection at the cyclopentane site.

Future work might show that the statistical distribution of angles between line segments provides information on the details of the in-plane defects.  The qualitative simulations presented here are only meant to suggest features of possible interest in experimental images.

\begin{figure}[tbp]
\includegraphics{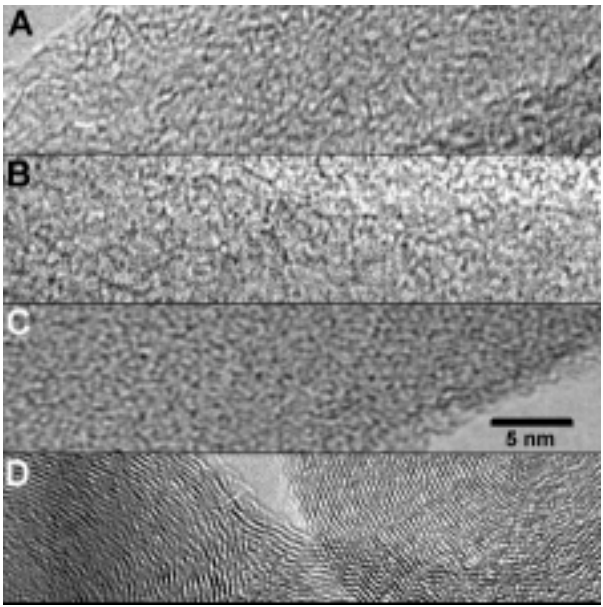}%
\caption{Experimental HREM images of four specimens: (A) core material from an onion in our negative 1914, (B) core material and (C) amorphous support material from a different onion in our negative 7005, and (D) terrestrial graphetized carbon.}
\label{Fig4}
\end{figure}

\begin{figure}[tbp]
\includegraphics{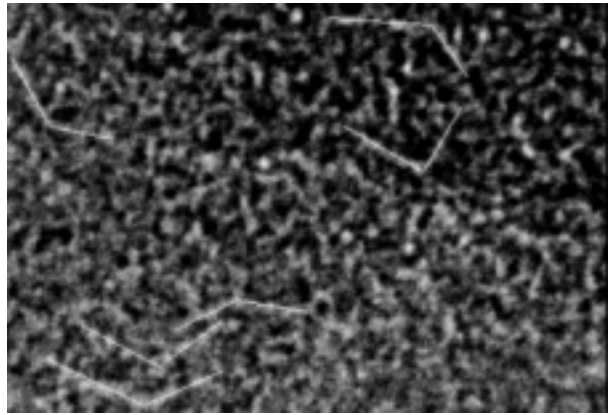}%
\caption{Inset of part of the 2nd core region above, in negative contrast with some intersecting linear features highlighted in white.}
\label{Fig5}
\end{figure}

Figure \ref{Fig4} shows experimental HREM images of four specimens.  The top two are of onion core material.  Power spectrum of these and/or adjacent image regions show strong 0.21 nm (100) rings, like those seen in core material electron diffraction and in the simulation of Fig \ref{Fig2b}.  Note in particular the non-parallel linear features in the top two images, not typical of the bottom two.  These features often manifest as an intersection between two line segments, as illustrated in the inset (from the core image B) in Fig \ref{Fig5}.  The line segments involved with this contrast are typically 2 to 5 nm in length.  Coherence widths suggested by diffraction are typically 4 nm \cite{Bernatowicz96}.  These observations are consistent, particularly since we expect our ability to recognize linear features in these images, in the presence of otherwise random superpositions, to be biased toward features larger than 2 nm.  

A caution is in order for those who wish to associate specific features in the experimental images with model structures.  Contrast depends on both specimen thickness and microscope defocus.  We've selected relatively sharp images, that don't show signs of global anisotropy e.g. due to drift or astigmatism.  However, detailed information on specimen thickness or defocus is not yet available.

\section{Summary}

We report simulations which suggest that the signature of 
a solid comprised of randomly-oriented 2 to 4 nm graphene sheets, in electron phase contrast images, will be images peppered with line segments due to those sheets which the beam encounters nearly edge-on, and with otherwise disordered contrast nonetheless showing power spectrum {\em rings} characteristic of the graphene 0.21nm spacing.  We further provide evidence, through simulations, that planar defects in these sheets will manifest as a pattern of line segment pairs intersecting at the sheet defect connecting them.  Lastly, we show that experimental HREM images of material found in the cores of interstellar graphite onions has a pattern of linear features consistent with these simulations.  These preliminary observations support the possibility that these presolar graphene cores manage three-dimensional growth atom-by-atom with help from sheet defects (e.g. cyclo-pentane or cyclo-heptane inserts randomly placed within sheets of predominantly hexagonal graphene rings).  

The observations also suggest two additional possibilities for future work.  (i) The initial growth of these cores might have taken place around a corannulene icospiral shell that is relatively ordered and less dense than the remainder of the core 
\cite{Kroto88}.  Careful HREM searches of the center of these onion cores might be able to identify it's presence.  (ii) Microscopy techniques which compensate for aberrations with the help of simulation, and thereby push the point resolution of the images down toward 0.1 nm, combined with further molecular modeling and image simulation of possible graphene sheet structures, might enable the kinds of analyses hinted at here to provide a more detailed picture of the structure and properties of this novel material, as well as of the processes involved in its formation.

\begin{acknowledgments}

  We would like to thank Kay Brewer, David Dawkins, Joyce Myers, Minh Truong and Dori Witt for their interest and work on this project, and the Enrico Fermi Institute's Roy Lewis and Washington University's Tom Bernatowicz for specimens and helpful discussions. 

\end{acknowledgments}

\bibliography{etmrefs}

\end{document}